\documentclass[12pt]{article}
\usepackage{enumerate}
\usepackage{amsfonts}
\usepackage{amsmath}
\usepackage{amssymb}
\usepackage{amsthm}

\def\H{\mathcal{H}}

\def\S{\mathfrak{S}}

\def\T{\mathfrak{T}}
\def\B{\mathfrak{B}}

\newcommand{\id}{\mathrm{Id}}
\newcommand{\Tr}{\mathrm{Tr}}

\newcommand{\shs}{\hspace{1pt}}

\newcounter{defin}  \newcounter{lemma}  \newcounter{theorem}
\newcounter{property} \newcounter{corol}  \newcounter{remark} \newcounter{example}

\newenvironment{lemma}{\par\refstepcounter{lemma}
     \textbf{Lemma \thelemma.} }{\rm\par}
\newenvironment{theorem}{\par\refstepcounter{theorem}
     \textbf{Theorem \thetheorem.}\ }{\rm\par}
\newenvironment{property}{\par\refstepcounter{property}
     \textbf{Proposition \theproperty.}\ }{\rm\par}
\newenvironment{corollary}{\par\refstepcounter{corol}
     \textbf{Corollary \thecorol.} }{\rm\par}
\newenvironment{definition}{\par\refstepcounter{defin}
     \textbf{Definition \thedefin.}\ }{\rm\par}
\newenvironment{remark}{\par\refstepcounter{remark}
     \textbf{Remark \theremark.}}{\rm\par}

\begin{document}

\title{Strong$^*$ convergence of quantum channels}

\author{M.E.~Shirokov,\footnote{email:msh@mi.ras.ru} \\\\
Steklov Mathematical Institute, Moscow, Russia}
\date{}
\maketitle

\begin{abstract}
In [arXiv:1712.03219] the existence of a strongly (pointwise) converging sequence of quantum channels that can not be represented as a reduction of
a sequence of unitary channels strongly converging to a unitary channel is shown. In this work we give a simple characterization of  sequences of quantum channels that have the above representation. The corresponding convergence is called the strong$^*$ convergence, since it is related to the convergence of selective Stinespring isometries of quantum channels in the strong$^*$ operator topology.

Some properties of the strong$^*$ convergence of quantum channels are considered. It is shown that for Bosonic Gaussian channels the strong$^*$ convergence coincides with the strong convergence.
\end{abstract}

\tableofcontents

\pagebreak

\section{Introduction}

The Stinespring theorem implies that any quantum channel $\Phi$ from a system $A$ to a system $B$ can be represented as
\begin{equation}\label{S-r}
\Phi(\rho)=\Tr_E V_{\Phi}\rho V^*_{\Phi},
\end{equation}
where $V_{\Phi}$ is an isometrical embedding of the input Hilbert space $\H_A$ into the tensor product of the output Hilbert  space $\H_B$ and some
Hilbert  space $\H_E$ typically called environment \cite{H-SCI,Wilde}. By using the Stinespring representation (\ref{S-r}) for any quantum channel $\Phi$ from $A$ to $B$ one can find such quantum systems $D$ and $E'$ that
\begin{equation}\label{ud-2}
  \Phi(\rho)=\Tr_{E'}\shs U_{\Phi\shs} \rho\otimes\sigma_{0} U_{\Phi}^*,
\end{equation}
where $\sigma_{0}$ is a pure state in $\S(\H_{D})$ and $U_{\Phi}$ is an unitary operator from $\H_{AD}$ onto $\H_{BE'}$ \cite{H-SCI}.
In the case $A=B$ one can take $D=E'=E$ in (\ref{ud-2}) \cite{Kraus,Wilde}.

Representation (\ref{ud-2}) called \emph{unitary dilation} of a quantum channel $\Phi$  allows to consider this channel as a reduction of some  unitary (reversible) channel between larger quantum systems \cite{H-SCI,Kraus,Wilde}.

In study of quantum channels and their information characteristics it is necessary to consider topology (convergence) on the set of all quantum channels between given quantum systems. In finite-dimensions the uniform convergence denerated by the diamond-norm metric  between quantum channels is widely used \cite{Kit,Paul},\cite[Ch.9]{Wilde}. But this convergence is \emph{too strong} for analysis of real variations of infinite-dimensional  channels \cite{SCT,W-EBN}. In this case it is natural to use the \emph{strong convergence} of  quantum channels generated  by the family of seminorms $\Phi\mapsto\|\Phi(\rho)\|_1$, $\rho\in\S(\H_A)$ \cite{AQC,Wilde+}. The strong convergence of a sequence $\{\Phi_n\}$ of channels to a channel $\Phi_0$  means that
\begin{equation}\label{star+}
\lim_{n\rightarrow\infty}\Phi_n(\rho)=\Phi_0(\rho)\,\textup{ for all }\rho\in\S(\H_A).
\end{equation}

It is easy to see that a sequence of unitary channels $\rho\mapsto U_n\rho U^*_n$ strongly converges to a
unitary channel $\rho\mapsto U_0\rho U^*_0$ if and only if the sequence $\{U_n\}$ converges to the operator $U_0$ in the strong operator topology.\footnote{The strong operator topology on the set of unitary operators  coincides with the weak, $\sigma$-weak and $\sigma$-strong operator topologies \cite{B&R}.} A characterization of the strong convergence of arbitrary quantum channels is presented in \cite[Theorem 1]{CSR}, it states that a sequence  $\{\Phi_n\}$ of quantum channels strongly converges to a quantum channel $\Phi_0$ if and only if  there is a quantum system $E$ and a sequence  $\{V_{\Phi_n}\}$ of isometries from $\H_A$ into $\H_{BE}$ converging to an isometry $V_{\Phi_0}$ in the strong operator topology such that (\ref{S-r}) holds for all $n$. \smallskip

The nontrivial part of this  characterization can be treated as continuity of the multi-valued map  $\Phi\mapsto V_{\Phi}$  (where $V_{\Phi}$ is the isometry from representation (\ref{S-r})) w.r.t. the strong convergence topology on the set of channels and the strong operator topology on the set of isometries. It turns out  that this continuity  does not imply continuity of the multi-valued map  $\Phi\mapsto U_{\Phi}$ (where $U_{\Phi}$ is the unitary from representation (\ref{ud-2})) w.r.t. these topologies. This means the existence of a sequence  $\{\Phi_n\}$ of quantum channels strongly converging to a quantum channel $\Phi_0$ that can not be represented in form (\ref{ud-2}) with some sequence  $\{U_{\Phi_n}\}$ of unitaries converging to a unitary operator $U_{\Phi_0}$ in the strong operator topology \cite{CSR}.

The above discontinuity of the unitary dilation is a specific feature of the strong convergence: by using the arguments from the proof of Theorem 1 in \cite{Kr&W} one can show that any sequence $\{\Phi_n\}$ of quantum channels converging to a quantum channel $\Phi_0$ w.r.t. the diamond norm \emph{can be represented} in form (\ref{ud-2}) with some sequence  $\{U_{\Phi_n}\}$ of unitaries converging to a unitary operator $U_{\Phi_0}$ in the operator norm topology \cite[Proposition 4]{CSR}.

The discontinuity of the unitary dilation with respect to the strong convergence of quantum channels  means  the existence of a strongly converging sequence of channels that \emph{has  no physical sense} within  the standard interpretation of a channel as a reduced unitary evolution of some larger system. Mathematically, this means that the strong convergence of quantum channels is \emph{too weak} for describing physical perturbations of quantum channels.\footnote{It seems reasonable to assume that all physical perturbations of a unitary channel $\rho\mapsto U\rho U^*$ is properly described by continuous deformations of the unitary $U$ in the strong operator topology (coinciding in this case with the weak, $\sigma$-weak and $\sigma$-strong operator topologies \cite{B&R}). I would be grateful for any comments concerning this question.}

The aim of this note is to describe all sequences of quantum channels that  can be represented as a reduction of
a sequence of unitary channels strongly converging to a unitary channel. We call the corresponding convergence of quantum channels \emph{the strong$\shs^*$ convergence}, since it is related to the convergence of selective Stinespring isometries of quantum channels in the strong$^*$ operator topology \cite{B&R}.

\bigskip

\section{Preliminaries}

Let $\mathcal{H}$ be a separable infinite-dimensional Hilbert space,
$\mathfrak{B}(\mathcal{H})$ the algebra of all bounded operators  on $\mathcal{H}$ with the operator norm $\|\cdot\|$ and $\mathfrak{T}( \mathcal{H})$ the
Banach space of all trace-class
operators on $\mathcal{H}$ with the trace norm $\|\!\cdot\!\|_1$. Let
$\mathfrak{S}(\mathcal{H})$ be  the set of quantum states (positive operators
in $\mathfrak{T}(\mathcal{H})$ with unit trace) \cite{H-SCI,Wilde}.


A \emph{quantum channel} $\,\Phi$ from a system $A$ to a system
$B$ is a completely positive trace preserving linear map from
$\mathfrak{T}(\mathcal{H}_A)$ into $\mathfrak{T}(\mathcal{H}_B)$ \cite{H-SCI,Wilde}. For any  quantum channel $\,\Phi:A\rightarrow B\,$ the Stinespring theorem (cf. \cite{St}) implies existence of a Hilbert space
$\mathcal{H}_E$ and of an isometry
$V_{\Phi}:\mathcal{H}_A\rightarrow\mathcal{H}_B\otimes\mathcal{H}_E$ such
that representation (\ref{S-r}) holds. The quantum  channel
\begin{equation}\label{c-channel}
\mathfrak{T}(\mathcal{H}_A)\ni
\rho\mapsto\widehat{\Phi}(\rho)=\mathrm{Tr}_{B}V_{\Phi}\rho
V_{\Phi}^{*}\in\mathfrak{T}(\mathcal{H}_E)
\end{equation}
is called \emph{complementary} to the channel $\Phi$
\cite[Ch.6]{H-SCI}. The complementary channel is uniquely defined up to \emph{isometrical equivalence}, i.e.
if $\,\widehat{\Phi}':A\rightarrow E'$ is the channel defined by formula  (\ref{c-channel})
via some other Stinespring isometry  $\,V'_{\Phi}:\H_A\rightarrow\H_B\otimes\H_{E'}\,$ then there exists a partial isometry
$W:\H_E\rightarrow\H_{E'}$ such that $\widehat{\Phi}'(\rho)=W\widehat{\Phi}(\rho)W^*$ and $\widehat{\Phi}(\rho)=W^*\widehat{\Phi}'(\rho)W$ for all $\rho\in \S(\H_A)$ \cite{H-c-ch}.\smallskip

The \emph{strong convergence} of  quantum channels is generated  by the family of seminorms $\Phi\mapsto\|\Phi(\rho)\|_1$, $\rho\in\S(\H_A)$ \cite{AQC}. The strong convergence of a sequence $\{\Phi_n\}$ of channels to a channel $\Phi_0$  means that
\begin{equation}\label{star++}
\lim_{n\rightarrow\infty}\Phi_n(\rho)=\Phi_0(\rho)\,\textup{ for all }\rho\in\S(\H_A).
\end{equation}
This convergence is more relevant for analysis of infinite-dimensional quantum channels than the diamond-norm convergence \cite{SCT,W-EBN}. Equivalent definitions of the strong convergence and its  properties are described in \cite{Wilde+}.

\smallskip

If $\Phi$ is a linear bounded map from $\T(\H_A)$ to $\T(\H_B)$ then the map $\Phi^*:\B(\H_B)\rightarrow\B(\H_A)$ defined by the relation
\begin{equation}\label{dual-r}
\Tr\shs \Phi(\rho)B=\Tr\shs\Phi^*\!(B)\rho\;\textrm{ for all }\;B\in\B(\H_B)
\end{equation}
is called \emph{dual} to the map $\Phi$ \cite{B&R,R&S}. If $\Phi$ is a channel acting on quantum states, i.e. a channel in the Schrodinger pucture, then $\Phi^*$ is a channel acting on quantum observables, i.e. a channel in the Heisenberg picture \cite{H-SCI,Wilde}.

The result in \cite{D-A} implies that the trace-norm convergence in (\ref{star++}) is equivalent to the convergence of the sequence $\{\Phi_n(\rho)\}$ to the state $\Phi_0(\rho)$ in the weak operator topology. So, by noting that the set $\S(\H_A)$ in (\ref{star++}) can be replaced by its subset consisting of pure states it is easy to show that \emph{the strong convergence of a sequence $\{\Phi_n\}$ of quantum channels to a channel $\Phi_0$  means, in the Heisenberg picture, that}
\begin{equation}\label{sc-ed}
w\shs\textup{-}\lim\limits_{n\rightarrow\infty}\Phi_n^*(B)=\Phi_0^*(B)\;\textit{ for all }\;B\in\B(\H_B),
\end{equation}
where $w\shs\textup{-}\lim$ denotes the limit in the weak operator topology in $\B(\H_A)$.

\section{On sequences of quantum channels  having  strongly converging unitary dilations}

The following theorem gives several criteria for existence of a strongly converging unitary dilation for a sequence of quantum channels.\smallskip

\begin{theorem}\label{ud-cr} \emph{Let $\,\{\Phi_n\}_{n\geq 0}$ be a sequence of quantum channels from $A$ to $B$. The following properties $(i)\textrm{-}\,(v)$ are equivalent:}
\begin{enumerate}[(i)]
   \item \emph{there exist quantum systems $D$ and $E$, a sequence $\{U_n\}_{n\geq0}$ of unitary operators from  $\H_{AD}$ onto $\H_{BE}$ and a sequence $\{\sigma_n\}$ of states  in $\,\S(\H_{D})$ converging to a state $\sigma_0$  such that $\Phi_n(\rho)=\Tr_{E}\shs U_n\rho\otimes\sigma_n U^*_n$ for all $\,n\geq0$ and $s\shs\textup{-}\lim\limits_{n\rightarrow\infty} U_n=U_0$;}\footnote{$\,s\shs\textup{-}\!\lim\limits_{n\rightarrow\infty} X_n=X_0$ denotes the strong convergence of a sequence $\{X_n\}$ to operator $X_0$.}
   \item \emph{there exist quantum systems $D$ and $E$, a sequence $\{U_n\}_{n\geq0}$ of unitary operators from  $\H_{AD}$ onto $\H_{BE}$ and a pure state $\sigma_0$ in $\S(\H_{D})$ such that $\Phi_n(\rho)=\Tr_{E}\shs U_n\rho\otimes\sigma_0 U^*_n$ for all $\,n\geq0$ and $\,s\shs\textup{-}\!\lim\limits_{n\rightarrow\infty} U_n=U_0$;}
   \item \emph{there exist a quantum system $E$ and a sequence $\{V_n\}_{n\geq0}$  of isometries from $\,\H_{A}$ into $\H_{BE}$ such that $\Phi_n(\rho)=\Tr_E V_n\rho V^*_n$ for all $\,n\geq0$, $\,s\shs\textup{-}\lim\limits_{n\rightarrow\infty} V_n=V_0$ and $\,s\shs\textup{-}\lim\limits_{n\rightarrow\infty} V^*_n=V^*_0$;}
  \item \emph{there exist a set of sequences $\{A^n_i\}_{n\geq0}$, $i\in I$, of operators from $\H_{A}$ to $\H_{B}$ such that $\Phi_n(\rho)=\sum_{i\in I} A^n_i\rho [A^n_i]^*$ for all $n\geq0$, $\,s\shs\textup{-}\lim\limits_{n\rightarrow\infty} A^n_i=A^0_i$ and $\,s\shs\textup{-}\lim\limits_{n\rightarrow\infty} [A^n_i]^*=[A^0_i]^*$ for each $\,i\in I$;}
\item \emph{$s\shs\textup{-}\lim\limits_{n\rightarrow\infty}\Phi_n^*(B)=\Phi_0^*(B)$ for all $\,B\in\B(\H_B)$.}\footnote{$\Phi^*$ is the dual map to the channel $\Phi$ defined by relation (\ref{dual-r}).}
\end{enumerate}
\smallskip
\emph{The equivalent properties $(i)\textrm{-}\,(v)$ imply  the strong convergence of the sequence $\{\Phi_n\}$ to the channel $\,\Phi_0$ (but the converse implication is not valid).}

\end{theorem}\medskip

\begin{remark}\label{ud-cr-r}
Property $\rm(v)$ in Theorem \ref{ud-cr} can be replaced by the following property, which is more easily verified in some cases\footnote{see the proof of Proposition \ref{Gauss} in Section 5.}
\begin{enumerate}
\item [$(\rm v')$]\emph{the sequence $\,\{\Phi_n\}$ strongly converges to the channel $\,\Phi_0$ and
$$
s\shs\textup{-}\lim\limits_{n\rightarrow\infty}\Phi_n^*(B)=\Phi_0^*(B)\quad\textrm{for all}\quad B\in\B_0,
$$
where $\,\B_0$ is a dense subset of $\,\B(\H_B)$ in the strong operator topology.}
\end{enumerate}
This follows from the proof of the implication $\rm (v)\Rightarrow(iii)$
in Theorem 1.
\end{remark}\smallskip

\emph{Proof of Theorem 1.} $\rm (i)\Rightarrow(iii).$ Since for any converging sequence of states in $\S(\H_{D})$ there is a converging sequence of purifications in
$\S(\H_{DR})$, where $R$ is some system, and $\,s\shs\textup{-}\lim_{n\rightarrow\infty} U_n=U_0$ implies $\,s\shs\textup{-}\lim_{n\rightarrow\infty} U_n\otimes I_R=U_0\otimes I_R$, we may assume that the sequence $\{\sigma_n\}$ consists of pure states. Let $\{\tau_n\}$ be a sequence of unit vectors in $\H_{D}$ converging to a unit vector $\tau_0$ such that $\sigma_n=|\tau_n\rangle\langle\tau_n|$ for all $n\geq 0$. For each $n$ let $V_n:\H_A\rightarrow\H_{BE}$ and $P_n:\H_A\rightarrow\H_{AD}$ be the operators defined by settings  $V_n|\varphi\rangle=U_n|\varphi\otimes\tau_n\rangle $ and $P_n|\varphi\rangle=|\varphi\otimes\tau_n\rangle$ for any $\varphi\in\H_A$. Then $\Phi_n(\rho)=\Tr_{E} V_n\rho V^*_n$ for all $n\geq 0$ and $s\shs\textup{-}\lim_{n\rightarrow\infty}V_n=V_0$. Since
$$
V_n^*|\psi\rangle \otimes |\tau_n\rangle=P_nV_n^*|\psi\rangle=[P_nV_n^*U_n]U_n^*|\psi\rangle
$$
for any vector $\psi$ in $\H_{BE}$, to show that $s\shs\textup{-}\lim_{n\rightarrow\infty} V^*_n=V^*_0$ it suffices to note that $s\shs\textup{-}\lim_{n\rightarrow\infty}U^*_n=U^*_0$ and that the operator $P_nV^*_nU_n$ is the orthogonal projector on the subspace $\H_A\otimes \{c\tau_n\}$ of $\H_{AD}$ for each  $n\geq 0$.\smallskip

$\rm (iii)\Rightarrow(iv).$ Let $\{\tau_i\}_{i\in I}$ be a basic in $\H_E$. For given $n$ and $i$ let $A^n_i$ be the operator from $\H_A$ to $\H_B$ such that $\langle\psi|A^n_i|\varphi\rangle=\langle\psi\otimes \tau_i|V_n|\varphi\rangle$ for any $\varphi\in\H_A$ and $\psi\in\H_B$. Then $\Phi_n(\rho)=\sum_{i\in I} A^n_i\rho [A^n_i]^*$ for all $n\geq0$. By noting that $V_n|\varphi\rangle=\sum_{i\in I} A^n_i|\varphi\rangle\otimes|\tau_i\rangle$ and $V_n^*|\varphi\otimes\tau_i\rangle=[A^n_i]^*|\varphi\rangle$  for any $\,i\,$ and $\varphi\in\H_A$ it is easy to show that
$\,s\shs\textup{-}\lim_{n\rightarrow\infty} A^n_i=A^0_i$ and $\,s\shs\textup{-}\lim_{n\rightarrow\infty} [A^n_i]^*=[A^0_i]^*$ for each $i\in I$.
\smallskip

$\rm (iv)\Rightarrow(iii).$ Let $V_n|\varphi\rangle=\sum_{i\in I} A^n_i|\varphi\rangle\otimes|\tau_i\rangle$ for any $\varphi\in \H_A$, where $\{\tau_i\}_{i\in I}$ is a basic in appropriate Hilbert space $\H_E$. Then $\Phi_n(\rho)=\Tr_E V_n\rho V^*_n$ for all $n\geq0$. Since $\,s\shs\textup{-}\lim_{n\rightarrow\infty} A^n_i=A^0_i$ for all $i\in I$, the sequence $\{V_n|\varphi\rangle\}$ weakly converges to the vector $V_0|\varphi\rangle$. The norm convergence of this sequence follows from the fact that all the operators $V_n$ are isometries. Since $\,s\shs\textup{-}\lim_{n\rightarrow\infty} [A^n_i]^*=[A^0_i]^*$ and $V_n^*|\varphi\otimes\tau_i\rangle=[A^n_i]^*|\varphi\rangle$ for all $i\in I$ and $n\geq0$, the sequence $\{V_n^*\}$ strongly converges to the operator $V_0^*$.\smallskip

$\rm (iii)\Rightarrow(ii).$ Let $C$ and $D$ be any infinite-dimensional quantum systems and $\sigma_0=|\tau_0\rangle\langle\tau_0|$, where $\tau_0$ is any unit vector in $\H_D$. If we identify the space $\H_A$ with the subspace $\H_A\otimes \{c\tau_0\}$ of $\H_{AD}$, then $\{V_n\}$ is a sequence of partial isometries from  $\H_{AD}$ to $\H_{BEC}\cong\H_{AD}$ strongly converging to the partial isometry $V_0$ such that $V^*_nV_n=V^*_0V_0$,  $\dim\ker V^*_nV_n=\dim\ker V_nV^*_n=+\infty$ for all $n\geq 0$ and $\,s\shs\textup{-}\lim_{n\rightarrow\infty} V_n^*=V_0^*$. So, the existence of the sequence $\{U_n\}$ with the required properties (with the system $EC$ in the role of $E$) follows from Proposition \ref{udc} in the Appendix. \smallskip

The implication $\rm (i)\Rightarrow(v)$ is stated in Proposition 3 in \cite{CSR}. Note that $\rm (v)$  directly follows from $\rm (iii)$, since $\Phi_n^*(B)=V_n^*B\otimes I_E V_n$ for any $B\in\B(\H_B)$ and all $n\geq 0$. The implication $\rm (ii)\Rightarrow(i)$ is trivial.\smallskip

$\rm (v)\Rightarrow(iii).$ Note that $\rm(v)$ imply  the strong convergence of the sequence $\{\Phi_n\}$ to the channel $\,\Phi_0$, since this convergence is equivalent to (\ref{sc-ed}). Assume first that the channel $\Phi_0$ has infinite Choi rank and $\,\Phi_0(\rho)=\Tr_E V_0\rho V^*_0$ is the \emph{minimal} Stinespring representation of the channel $\Phi_0$ \cite[Ch.6]{H-SCI}. It means that the set
\begin{equation}\label{theset}
\left\{[B\otimes I_E] V_0|\varphi\rangle\shs|\shs B\in\B(\H_B),\varphi\in\H_A\right\}
\end{equation}
is dense in $\H_{BE}$. By Lemma \ref{ud-cr-l} below there is a sequence $\{V_n\}$ of isometries from $\H_A$ into $\H_{BE}$ strongly converging to $V_0$ such that $\Phi_n(\rho)=\Tr_E V_n\rho V^*_n$ for all $\,n$. Since $\Phi_n^*(B)=V_n^*B\otimes I_E V_n$ for any $B\in\B(\H_B)$, by using the condition  $\,s\shs\textup{-}\lim_{n\rightarrow\infty} V_n=V_0$
it is easy to show that $\,\lim_{n\rightarrow\infty} V^*_n|\psi\rangle =V^*_0|\psi\rangle$ for any vector $\psi$ from the set (\ref{theset}).

If the channel $\Phi_0$ has finite Choi rank then take any channel $\Psi: A\rightarrow C$ with infinite Choi rank and consider the sequence of  channels
$$
\widetilde{\Phi}_n(\rho)=\textstyle\frac{1}{2}{\Phi}_n(\rho)\oplus\textstyle\frac{1}{2}\Psi(\rho)
$$
from $\T(\H_A)$ into $\T(\H_B\oplus\H_C)$  strongly converging to the channel $\widetilde{\Phi}_0(\rho)=\textstyle\frac{1}{2}{\Phi}_0(\rho)\oplus\textstyle\frac{1}{2}\Psi(\rho)$ with infinite Choi rank.
It is easy to see that $\rm (v)$ implies
$$
s\shs\textup{-}\lim\limits_{n\rightarrow\infty}\widetilde{\Phi}_n^*(B)=\widetilde{\Phi}_0^*(B)\quad\forall B\in\B(\H_B\oplus\H_C).
$$
By the above part of the proof
property $\rm (iii)$ holds for the sequence $\{\widetilde{\Phi}_n\}$, i.e.
there exist a quantum system $E'$ and a sequence $\{\widetilde{V}_n\}_{n\geq0}$  of isometries from $\,\H_{A}$ into $(\H_{B}\oplus\H_{C})\otimes\H_{E'}$ such that $\widetilde{\Phi}_n(\rho)=\Tr_{E'} \widetilde{V}_n\rho \widetilde{V}^*_n$ for all\break $\,n\geq0$, $\,s\shs\textup{-}\lim\limits_{n\rightarrow\infty} \widetilde{V}_n=\widetilde{V}_0$ and $\,s\shs\textup{-}\lim\limits_{n\rightarrow\infty} \widetilde{V}^*_n=\widetilde{V}^*_0$. Let $P_B$ be the projector on the subspace $\H_B$ of $\H_B\oplus\H_C$. Then
$$
\Phi_n(\rho)=2P_B\widetilde{\Phi}_n(\rho)P_B=2\Tr_{E'} [P_B\otimes I_{E'}][ \widetilde{V}_n\rho \widetilde{V}^*_n  ][P_B\otimes I_{E'}]\quad \forall n\geq 0.
$$
Hence  $V_n=\sqrt{2}[P_B\otimes I_{E'}]\widetilde{V}_n$ is a Stinespring isometry for the channel $\Phi_n$ for all $n\geq 0$. Since
$\,s\shs\textup{-}\lim\limits_{n\rightarrow\infty} V_n=V_0$ and $\,s\shs\textup{-}\lim\limits_{n\rightarrow\infty} V^*_n=V^*_0$, property $\rm(iii)$  holds for the sequence $\{\Phi_n\}$.\smallskip

The last assertion of the theorem follows from the equivalent definition (\ref{sc-ed}) of the strong convergence and Corollary 3 in \cite{CSR}. $\square$\medskip

\begin{lemma}\label{ud-cr-l} \emph{Let $\,\{\Phi_n\}$ be a sequence of quantum channels from $A$ to $B$ strongly converging to a channel $\Phi_0$ with infinite Choi rank. For any given Stinespring isometry $V_0:\H_A\rightarrow \H_{BE}$ of the channel $\,\Phi_0$ there is a sequence $\{V_n\}$ of isometries from $\H_A$ into $\H_{BE}$ strongly converging to $V_0$ such that $\Phi_n(\rho)=\Tr_E V_n\rho V^*_n$ for all $\,n$.}
\end{lemma}\smallskip

\emph{Proof.} Since $\H_E$ is an infinite-dimensional Hilbert space, for each $n$ there exists an isometry $V_n:\H_A\rightarrow \H_{BE}$ such that
$\Phi_n(\rho)=\Tr_E V_n\rho V^*_n$. By the proofs of Lemma 1 and Theorem 1 in \cite{CSR} (based on the results from \cite{Kr&W}) there is a sequence $\{C_n\}$ of contractions in $\B(\H_E)$ such that
the sequence of isometries
$$
\widehat{V}_n=[I_B\otimes C_n]V_n\oplus[I_B\otimes\sqrt{I_E-C^*_nC_n}]V_n
$$
from $\H_A$ into $\H_B\otimes(\H_E\oplus\H_E)$ strongly converges to the isometry $V_0\oplus 0$. By simple continuity arguments we may assume that all the operators $C_n$ are non-degenerate. Let $U_n$ be the isometry from the  polar decomposition of $C_n$. Since the sequences $\,[I_B\otimes C_n]V_n$ and  $\,[I_B\otimes\sqrt{I_E-C^*_nC_n}]V_n$ strongly converge to the isometry $V_0$ and to the zero operator correspondingly, it is easy to show that the sequence  $W_n=(I_B\otimes U_n)V_n$ strongly converges to the isometry $V_0$. It is clear that $\Phi_n(\rho)=\Tr_E W_n\rho W^*_n$ for all $\,n$. $\square$
\medskip

According to the operator theory terminology a sequence $\{T_n\}$ of operators from $\H_A$ into $\H_B$ is called \emph{strongly$^*$  converging} to an operator $T_0$ if $\,s\shs\textup{-}\lim_{n\rightarrow\infty} T_n=T_0\,$ and $\,s\shs\textup{-}\lim_{n\rightarrow\infty} T^*_n=T^*_0\,$ \cite{B&R}. So, Theorem \ref{ud-cr} states, in particular, that the existence of strongly converging sequence of unitary dilations is equivalent to the existence of strongly$^*$  converging
sequence of Stinespring isometries. This motivates the following\smallskip

\begin{definition}\label{star-s-c}
A sequence $\,\{\Phi_n\}$ of quantum channels is called \emph{strongly$^*$  converging to a channel $\,\Phi_0$} if the equivalent properties $\rm (i)\shs\textrm{-}(v)$ in Theorem \ref{ud-cr} hold.
\end{definition}\smallskip

Corollary 3 in \cite{CSR} implies that the strong$^*$  convergence of quantum channels is stronger than
the strong convergence. The difference between these types of convergence is best seen on the Heisenberg picture:
as mentioned at the end of Section 2 the strong convergence of a sequence $\{\Phi_n\}$ to a channel $\Phi$ can be defined as
$$
w\shs\textup{-}\lim\limits_{n\rightarrow\infty}\Phi_n^*(B)=\Phi_0^*(B)\;\textup{ for all }\;B\in\B(\H_B),
$$
(the limit in the weak operator topology) while the strong$^*$  convergence of this sequence means that
\begin{equation}\label{s-s-c-d}
s\shs\textup{-}\lim\limits_{n\rightarrow\infty}\Phi_n^*(B)=\Phi_0^*(B)\;\textup{ for all }\;B\in\B(\H_B).
\end{equation}

The below example shows, in particular, that the strong$^*$  convergence is substantially
weaker than the uniform (diamond norm) convergence. \smallskip

\textbf{Example.} Let $\Phi$ be an arbitrary quantum channel from $A$ to $B$ and  $\{P_n\}$ a sequence of finite rank projectors in $\B(\H_B)$ strongly converging to the unit operator $I_{\H_B}$. Let
$$
\Phi_n(\rho)=P_n \Phi(\rho) P_n+[\Tr(I_{\H_B}-P_n)\Phi(\rho)\shs]\sigma
$$
for all $\,n$, where $\sigma$ is a given state in $\S(\H_B)$. It is clear that the sequence $\{\Phi_n\}$ strongly converges
to the channel $\Phi$, but  it does not converge uniformly
to $\Phi$ in general (it suffices to consider the case $A=B$, $\Phi=\id_{\H_A}$). Sequences of this type  are used in \cite{AQC} for approximation of infinite-dimensional quantum channels by channels with finite-dimensional output system. Since the map $B\mapsto \Phi^*(B)$ is continuous w.r.t. the strong operator topology, we have
$$
s\shs\textup{-}\lim\limits_{n\rightarrow\infty}\Phi^*_n(B)=s\shs\textup{-}\lim\limits_{n\rightarrow\infty}\Phi^*(P_n B P_n+[I_{\H_B}-P_n]\Tr B\sigma)=\Phi^*(B)\quad \forall B\in\B(\H_B).
$$
So, the sequence $\{\Phi_n\}$ strongly$^*$  converges
to the channel $\Phi$.
\smallskip

By using (\ref{s-s-c-d}) as the simplest definition of the strong$^*$  convergence
it is easy to show that this convergence is preserved under  basic manipulations with quantum channels.\footnote{Similar statements for the strong convergence are proved explicitly in \cite{Wilde+}.}\smallskip

\begin{corollary}\label{ud-cr-c-1} \emph{Let $\,\{\Phi_n\}$ and $\,\{\Psi_n\}$ be  sequences of quantum channels from $A$ to $B$ and from $C$ to $D$
correspondingly that strongly$\shs^{*}$
converge to quantum channels $\,\Phi_0$ and  $\,\Psi_0$.}\smallskip

A) \emph{The sequence $\{\Phi_n\otimes\Psi_n\}$ of channels from $AC$ to $BD$ strongly$\shs^*$
converges to the channel $\,\Phi_0\otimes\Psi_0$.}\smallskip

B) \emph{If $B=C$ then the sequence $\{\Psi_n\circ\Phi_n\}$ of channels from $A$ to $D$ strongly$\shs^*$
converges to the channel $\,\Psi_0\circ\Phi_0$.}
\end{corollary}\smallskip

By using property $\rm(iii)$ in Theorem \ref{ud-cr} as a criterion of the strong$^*$  convergence it is easy to prove  the following observation which shows the \emph{continuity of the complementary operation}
$\,\Phi\mapsto\widehat{\Phi}\,$ (defined in (\ref{c-channel})) with respect to the strong$^*$  convergence of quantum channels.\smallskip

\begin{corollary}\label{ud-cr-c-1} \emph{If $\,\{\Phi_n\}$ is a sequence of quantum channels from $A$ to $B$ strongly$\shs^*$  converging to a channel $\,\Phi_0$ then there exists a sequence $\,\{\Psi_n\}$ of quantum channels  from $A$ to some system $E$ strongly$\shs^*$  converging  to a channel $\,\Psi_0$ from $A$ to $E$ such that $\,\Psi_n=\widehat{\Phi}_n$ for all $\,n\geq0$.}
\end{corollary}\vspace{-5pt}

\section{Representation of converging channels via a partial trace channel}

The Stinespring representation (\ref{S-r}) of any quantum channel $\Phi_0$ from $A$ to $B$ means that
$\Phi_0(\rho)=\Theta(V_0\rho V^*_0)$, where $\Theta(\rho)=\Tr_E \rho$ is the partial trace channel from $BE$ to $B$
and $V_0$ is an isometrical embedding of $\H_A$ into $\H_{BE}$.
Assume that $\{W_n\}$ is a sequence of partial isometries on $\H_{BE}$ such that $W_n^*W_n=P_0$ for all $n$, where $P_0$ is the projector
on the subspace $V_0(\H_A)$. Let
\begin{equation}\label{p-t-c}
\Phi_n(\rho)=\Theta(W_nV_0\rho V^*_0W_n^*),\quad \rho\in\S(\H_A),
\end{equation}
be a quantum channel from $A$ to $B$ for each $n$. It is easy to show that
\begin{itemize}
 \item if the sequence $\{W_n\}$ converges to the projector $P_0$ in the operator norm then the sequence $\{\Phi_n\}$ converges to the channel $\Phi_0$ in the diamond norm;
 \item if the sequence $\{W_n\}$ strongly converges to the projector $P_0$ then the sequence $\{\Phi_n\}$ strongly converges to the channel $\Phi_0$;
 \item if the sequence $\{W_n\}$ strongly$^*$ converges\footnote{It means that $\,s\shs\shs\textup{-}\!\lim\limits_{n\rightarrow\infty} W_n=P_0$ and $\,s\shs\textup{-}\!\lim\limits_{n\rightarrow\infty} W^*_n=P_0$ \cite{B&R}.} to the projector $P_0$ then the sequence $\{\Phi_n\}$ strongly$^*$ converges to the channel $\Phi_0$  (see Def.1).
\end{itemize}

The results in \cite{Kr&W},\cite{CSR} and Theorem 1 in Section 3 imply \smallskip
\begin{property}\label{ptr}
\emph{Any sequence $\{\Phi_n\}$ of channels converging in the diamond norm (correspondingly, strongly converging, strongly$\shs^*$ converging) to a channel $\Phi_0$ can be represented in the form (\ref{p-t-c}) with some isometrical embedding $V_0$ of $\H_A$ into $\H_{BE}$ and some sequence $\{W_n\}$ of partial isometries converging to the projector $P_0$ w.r.t the operator norm topology (correspondingly, the strong operator topology, the strong$\shs^*$  operator topology).} \smallskip
\end{property}

Usefulness of the  representation (\ref{p-t-c}) of strongly converging sequences of quantum channels is illustrated by the proof of Theorem 2 in \cite{CSR}.

\section{Convergence of Bosonic Gaussian channels}

In this section we show that the strong$^*$ and strong convergences  coincide
on the class Bosonic Gaussian channels playing a central role in continuous variable quantum information theory
\cite{H-SCI,W&C}.

Let $\mathcal{H}_{X}$ $(X=A,B,...)$ be the space of irreducible representation of
the Canonical Commutation Relations (CCR)
\begin{equation*}
W_X(z)W_X(z^{\prime })=e^{\shs\left[-\frac{i}{2}z^{\top}\!\Delta_{X}z^{\prime }\right]} W_X(z^{\prime }+z),\quad z,z'\in Z_X,  
\end{equation*}
with a symplectic space $(Z_{X},\Delta _{X})$ and the
Weyl operators $W_{X}(z)$ \cite[Ch.12]{H-SCI}. Denote by $s_X$ the number of modes of the system $X$, i.e. $2s_X=\dim Z_X$.\smallskip

A state $\rho$ in $\S(\H_X)$ is called Gaussian if it has Gaussian characteristic function
\begin{equation}\label{gaus-ch-f}
\phi_{\rho}(z)\doteq\Tr W_X(z)\rho=e^{\shs\left[ imz-\frac{1}{2}z^{\top}\!\sigma z\right]},
\end{equation}
where $m$ is a $\,2s_X$-dimensional real
row and $\sigma$ is a  $\,(2s_X)\times(2s_X)$ real symmetric matrix satisfying the inequality
\begin{equation}\label{cmc}
\sigma \geq \pm i\Delta _{X}.
\end{equation}
The row $m$ consists of the mean values of the canonical observables at the state $\rho$, while $\sigma$ is
the covariance matrix of these observables \cite[Ch.12]{H-SCI}.\smallskip

A Bosonic Gaussian channel  $\Phi:\mathfrak{T}(\mathcal{H}_{A})\rightarrow
\mathfrak{T}(\mathcal{H}_{B})$ is defined via the action of its dual
$\Phi^{\ast }:\mathfrak{B}(\mathcal{H}_{B})\rightarrow
\mathfrak{B}(\mathcal{H}_{A})$ on the Weyl operators:
\begin{equation}\label{blc}
\Phi^{\ast}(W_{B}(z))=W_A(Kz)e^{\shs\left[ i\ell z-\frac{1}{2} z^{\top}\!\alpha
z\right]},\quad z\in Z_B,
\end{equation}
where $K$ is a linear operator $Z_{B}\rightarrow Z_{A}$, $\ell\,$ is a $\,2s_B$-dimensional real
row and $\alpha\,$ is a  $\,(2s_B)\times(2s_B)$ real symmetric matrix satisfying the inequality
\begin{equation*}
\alpha \geq \pm \frac{i}{2}\left[ \Delta _{B}-K^{\top}\Delta_{A}K\right].
\end{equation*}

If $\rho$ is a state in $\S(\H_A)$ with the characteristic function $\,\phi_{\rho}(z)\doteq\Tr W_A(z)\rho\,$ and $\Phi$ is a Gaussian channel defined in (\ref{blc}) then the state $\Phi(\rho)$ has the characteristic function
\begin{equation}\label{out-cf-g}
\phi_{\Phi(\rho)}(z)=\Tr W_B(z)\Phi(\rho)=\Tr \Phi^*(W_B(z))\rho=\phi_{\rho}(Kz)e^{\shs\left[ i\ell z-\frac{1}{2} z^{\top}\!\alpha z\right]}.
\end{equation}
In particular, if $\rho$ is a Gaussian state with the characteristic function (\ref{gaus-ch-f}) then $\Phi(\rho)$ is a Gaussian state with the characteristic function
\begin{equation}\label{out-cf}
\phi_{\Phi(\rho)}(z)=e^{\shs\left[ i(mK +\ell)z-\frac{1}{2} z^{\top}\!(\alpha+K^{\top}\!\sigma K)z\right]}.
\end{equation}

\begin{property}\label{Gauss}
\emph{For Bosonic Gaussian channels  the strong$\shs^*$ convergence is equivalent to the
strong convergence and  weaker than the uniform (diamond norm) convergence.}
\end{property}\smallskip

\emph{Proof.} Let $\{\Phi_n\}$ be a sequence of Gaussian channels strongly converging to a Gaussian  channel $\Phi_0$.
By using Lemma \ref{Gauss-l} below and the coincidence of the weak operator topology with the
strong operator topology on the set of unitary operators it is easy to show that $\,s\shs\textup{-}\!\lim\limits_{n\rightarrow\infty}\Phi_n^{\ast}(W_{B}(z))=\Phi_0^{\ast}(W_{B}(z))$ for all $z\in Z_B$. Since
the linear hull of the family  $\{W_{B}(z)\}_{z\in Z_B}$ is dense in $\B(\H_B)$ in the strong operator topology,  the sequence $\{\Phi_n\}$  strongly$^*$ converges to the channel $\Phi_0$ by
Remark \ref{ud-cr-r} after Theorem \ref{ud-cr}.

To show that the strong$^*$ convergence is weaker than the uniform (diamond norm) convergence
one can consider  the single-mode Bosonic quantum limited attenuator defined by its action
on the family $\{|\eta\rangle\langle \eta|\}_{\eta\in \mathbb{C}}$ of coherent states as follows
$$
\Phi_{k}(|\eta\rangle\langle \eta|)=|k\eta\rangle\langle k\eta|.
$$
It is proved in \cite{W-EBN} that $\|\Phi_{k}-\Phi_{k'}\|_{\diamond}=2$ for all $k\neq k'$. It is also mentioned in \cite{W-EBN} that
the channel $\Phi_{k'}$ strongly (and hence strongly$^*$) converges to the channel $\Phi_{k}$ as $k'\rightarrow k$.
Other examples showing the nonequivalence of the strong$^*$ and uniform convergences of Gaussian channels can be found in \cite{Wilde+}. $\square$
\smallskip

\begin{lemma}\label{Gauss-l}
\emph{Let $\,\Phi_n$, $n\geq0$, be Gaussian channels between given Bosonic systems $A$ and $B$ defined by relation (\ref{blc})
with parameters $K_n, \ell_n, \alpha_n$. The sequence $\{\Phi_n\}$ strongly converges to the channel $\Phi_0$ if and only if
\begin{equation}\label{sc-gc}
  \lim_{n\rightarrow\infty} K_n=K_0,\quad\lim_{n\rightarrow\infty} \ell_n=\ell_0\quad\textit{and}\quad\lim_{n\rightarrow\infty} \alpha_n=\alpha_0,
\end{equation}
where the limits in any topology on sets of finite matrices (rows).}
\end{lemma}\smallskip

\emph{Proof.} Let $\rho$ be any state in $\S(\H_A)$. It follows from (\ref{out-cf-g}) that condition (\ref{sc-gc}) implies poinwise convergense of the sequence $\{\phi_{\Phi_n(\rho)}(z)\}$ to the function
$\phi_{\Phi_0(\rho)}(z)$. By the quantum version of Levy's continuity theorem (see \cite{QCT}) this
implies convergence of the sequence $\{\Phi_n(\rho)\}$ to the state $\Phi_0(\rho)$. \smallskip

Assume that the sequence $\{\Phi_n\}$ strongly converges to the channel $\Phi_0$. Then for any Gaussian state $\rho$ the sequence
$\{\phi_{\Phi_n(\rho)}(z)\}$ pointwise converges to the function $\phi_{\Phi_0(\rho)}(z)$. By expression (\ref{out-cf}) this means that
$$
\lim_{n\rightarrow\infty} e^{\shs\left[ i(mK_n+\ell_n)z-\frac{1}{2}z^{\top}\!(\alpha_n+K_n^{\top}\!\sigma K_n)z\right]}=e^{\shs\left[ i(mK_0+\ell_0)z-\frac{1}{2}z^{\top}\!(\alpha_0+K_0^{\top}\!\sigma K_0)z\right]}
$$
for all $z\in Z_B$. Since this relation holds for any mean row $m$ and covariance matrix $\sigma$ satisfying (\ref{cmc}), it is easy
show the validity of condition (\ref{sc-gc}). $\square$ \smallskip

Proposition \ref{Gauss} states that any  sequence of Gaussian channels
strongly converging to a Gaussian channel can be represented as a reduction of
a sequence of unitary channels strongly converging to a unitary channel.\smallskip

At the same time, it is well known that any Gaussian channel has a Gaussian unitary dilation, i.e. it
can be represented as a reduction of a Gaussian  unitary channel\footnote{Gaussian  unitary channel is a channel $\rho\mapsto U_T\rho U^*_T$, where $U_T$
is the canonical unitary corresponding to a symplectic trasformation $T$ \cite[Ch.12]{H-SCI}.}  between  composite Bosonic systems
\cite{G-D1,G-D2}\cite[Ch.12]{H-SCI}. So, it seems reasonable to assume that Proposition \ref{Gauss} can be strengthened as follows

\smallskip
\textbf{Conjecture.} \emph{Any sequence of Gaussian channels
strongly converging to a Gaussian channel can be represented as a reduction of
a sequence of Gaussian unitary channels strongly converging to a Gaussian unitary channel}.
\smallskip

This conjecture seems natural from the physical point of view but the explicit forms of Gaussian unitary dilations
of a single Gaussian channel presented in \cite{G-D1,G-D2} show that its direct proof requires serious technical efforts.

\section*{Appendix}

Below we present results concerning possibility to dilate a strongly converging sequence of partial isometries to
strongly converging sequence of unitary operators.\footnote{I am sure that these results can be found in the literature. So, I would be grateful for any references concerning this question.}

 \smallskip

\begin{property}\label{udc} \emph{Let $\,\{V_n\}$ be a sequence of partial isometries on a separable Hilbert space $\H$  strongly converges to a partial isometries $V_0$ such that $V_n^*V_n=V_0^*V_0=P$  and $\,\dim\mathrm{Ker}P=\dim\mathrm{Ker}Q_n\leq+\infty,\,$ where $Q_n=V_nV^*_n$,  for all $\,n\geq 0$.
The following properties are equivalent:}
\begin{enumerate}[(i)]
  \item \emph{there exists a sequence $\,\{U_n\}$ of unitaries on $\H$ strongly converging
to a unitary operator $U_0$ such that $\,U_nP=V_n$ for all $\,n\geq0$;}
  \item \emph{the sequence $\{Q_n\}$ strongly converges to the operator $Q_0$;}
  \item \emph{the sequence $\{V^*_n\}$ strongly converges to the operator $V^*_0$.}
\end{enumerate}
\end{property}\smallskip

\begin{remark}\label{r1}
A sequence $\,\{V_n\}$ of partial isometries satisfying
the assumptions of Proposition \ref{udc} for which the properties $\rm (i)\textrm{-}(iii)$  do not hold can be found in the proof of Corollary 3 in \cite{CSR}.
\end{remark}
\smallskip

\emph{Proof.} Since all the partial isometries have the same initial space, the sequence $\{W_n=V_nV_0^*\}$ consists of partial isometries and strongly converges to the projector $Q_0=V_0V_0^*$. Note that $W_nW_n^*=Q_n$ and $W^*_nW_n=Q_0$ for all $n$. So, the assertion of the proposition follows from Lemma \ref{udc-l} below. $\square$
\smallskip

\begin{lemma}\label{udc-l} \emph{Let $\,\{S_n=\{\varphi^n_i\}_{i\in I}\}_{n\geq0}$ be a sequence of orthonormal systems of vectors in a separable Hilbert space $\H$ such that $\,\dim S_n^{\bot}=\,\dim S_0^{\bot}\leq+\infty$ for all $\,n$. Let $\,P_n=\sum_{i\in I}|\varphi^n_i\rangle\langle\varphi^n_i|\,$ be the projector on the subspace $\H_n$ generated by $S_n$ and $\,W_n=\sum_{i\in I}|\varphi^n_i\rangle\langle\varphi^0_i|\,$  a partial isometry. Assume that $\,\lim_{n\rightarrow\infty} \varphi^n_i=\varphi^0_i$ for each $i\in I$. The following properties are equivalent:}
\begin{enumerate}[(i)]
  \item \emph{for each $\,n\geq 0\,$ there is an orthonormal basis $S^\mathrm{e}_n=\{\varphi^n_i\}_{i\in I}\cup\{\psi^n_j\}_{j\in J}$ in $\H$ obtained by extension of the system $S_n$ such that $\,\lim_{n\rightarrow\infty} \psi^n_j=\psi^0_j$ for each $j\in J$;}
  \item \emph{the sequence $\{P_n\}$ strongly converges to the operator $P_0$;}
  \item \emph{the sequence $\{W^*_n\}$ strongly converges to the operator $P_0$;}
\end{enumerate}
\end{lemma}\smallskip

Proof. $\rm(i)\Rightarrow(iii)$. It follows from $\rm(i)$ that
$$
U_n=\sum_{i\in I}|\varphi^n_i\rangle\langle\varphi^0_i|+\sum_{j\in J}|\psi^n_j\rangle\langle\psi^0_j|
$$
is an unitary operator strongly converging to the unit operator $I_{\H}$ as $n\rightarrow\infty$. Then the unitary operator $U^*_n$ strongly converges to the unit operator as well, i.e.
$$
\sum_{i\in I}|\varphi^0_i\rangle\langle\varphi^n_i|\theta\rangle\oplus\sum_{j\in J}|\psi^0_j\rangle\langle\psi^n_j|\theta\rangle\;\rightarrow\;
\sum_{i\in I}|\varphi^0_i\rangle\langle\varphi^0_i|\theta\rangle\oplus\sum_{j\in J}|\psi^0_j\rangle\langle\psi^0_j|\theta\rangle
$$
as $n\rightarrow\infty$ for any vector $\theta$ in $\H$. Hence $W^*_n$ strongly converges to $P_0$. \smallskip

$\rm(iii)\Rightarrow(ii)$. Since $W_n$ strongly converges to $P_0$  by the assumption, it follows from $\rm(iii)$ that  $P_n=W_nW_n^*$ strongly converges to $P_0$. \smallskip

$\rm(ii)\Rightarrow(i)$.  Let $S^\mathrm{e}_0=\{\varphi^0_i\}_{i\in I}\cup\{\psi^0_j\}_{j\in J}$ be an orthonormal basis (o.n.b. in what follows) in $\H$ obtained by extension of the system $S_0$. Sequentially applying Lemma \ref{udc-ll} below one can construct, for any natural $m$ and $n$, an orthonomal system $\{ \alpha^n_1,...,\alpha^n_m \}$ in $S_n^{\bot}$ in such a way that $\lim_{n\rightarrow\infty}\alpha^n_j=\psi^0_j$ for all $j=\overline{1,m}$. This gives the required sequence of o.n.b. $S^\mathrm{e}_n=S_n\cup\{\psi^n_j\}_{j\in J}$ in the case $\dim S_0^{\bot}<+\infty$. If $\dim S_0^{\bot}=+\infty$ this sequence can be constructed as follows:
$$
\begin{array}{l}
   \psi^1_1=\alpha^1_1\textrm{\, and \,}\{\psi^1_j\}_{j>1}\textrm{ is any o.n.b. in }[\{ \alpha^1_1\}\cup S_1]^{\bot},\\
   \psi^2_1=\alpha^2_1, \psi^2_2=\alpha^2_2\textrm{\, and\, }\{\psi^2_j\}_{j>2}\textrm{ is any o.n.b. in }[\{ \alpha^2_1,\alpha^2_2 \}\cup S_2]^{\bot},\\
   ..............................\\
 \psi^n_1=\alpha^n_1,..., \psi^n_n=\alpha^n_n\textrm{\, and\, }\{\psi^n_j\}_{j>n}\textrm{ is any o.n.b. in }[\{ \alpha^n_1,...,\alpha^n_n \}\cup S_n]^{\bot},\\
 ..............................
\end{array}
$$

\begin{remark}\label{r2} A sequence $\,\{S_n\}$ of orthonormal systems satisfying
the assumptions of Lemma \ref{udc-l} for which properties $\rm (i)\textrm{-}(iii)$ of this lemma do not hold can
be easily constructed: let $\{\tau_i\}$ be a countable orthonormal system of vectors,
$\varphi^n_i=\tau_i$ for all $i\neq n$ and $\varphi^n_n=\psi$, where $\psi$ is any unit vector in $\{\tau_i\}^{\bot}$.
\end{remark}\smallskip

\begin{lemma}\label{udc-ll} \emph{Let the assumptions of Lemma \ref{udc-l} hold and $\psi_0$ be any unit vector in $S_0^{\bot}$. If the sequence $\{P_n\}$ strongly converges to the operator $P_0$ then there is a sequence $\{\psi_n\}$ of unit vectors converging to the unit vector $\psi_0$ such that $\,\psi_n\in S_n^{\bot}$ for all $\,n$.}
\end{lemma}\smallskip

\emph{Proof.} Let $\bar{P}_n=I_{\H}-P_n$ and $|\psi_n\rangle=\bar{P}_n|\psi_0\rangle/\|\bar{P}_n|\psi_0\rangle\|$ if $\|\bar{P}_n|\psi_0\rangle\|\neq0$ and
$|\psi_n\rangle$ be any vector in $S_n^{\bot}$ otherwise. Since the sequence $\{\bar{P}_n\}$ strongly converges to the operator $\bar{P}_0$ and
$\bar{P}_0|\psi_0\rangle=|\psi_0\rangle$ the sequence $\{|\psi_n\rangle\}_n$ has the required properties. $\square$

\bigskip

I am grateful to Frederik vom Ende for the valuable communication.  I am
also grateful to A.S.Holevo, G.G.Amosov, A.V.Bulinski, V.Zh.Sakbaev, T.V.Shulman and M.M.Wilde for useful discussion. 

\end{document}